\documentclass{emulateapj}
\usepackage{amsmath,amsthm,amssymb}
\shorttitle{Plane-Parallel Relativistic Shock}
\shortauthors{Nakayama \& Shigeyama}

\begin{document}

\title{Self-Similar Evolution of Relativistic Shock Waves Emerging from Plane-Parallel  Atmospheres}

\author{Kazunori Nakayama}
\affil{Department of Astronomy, School of Science, University of Tokyo, Bunkyo-ku, 
        Tokyo 113-0033, Japan}
\and
\author{Toshikazu Shigeyama}
\affil{Research Center for the Early Universe, Graduate School of Science, 
        University of Tokyo, Bunkyo-ku, Tokyo 113-0033, Japan}

\begin{abstract}
We study the evolution of  the ultra-relativistic shock wave in  a plane-parallel atmosphere adjacent to a vacuum and the subsequent breakout phenomenon. When the density distribution has a power law with the distance from the surface, there is a self-similar motion of the fluid 
before and after the shock emergence. The time evolution of the Lorentz factor of the shock front is assumed to follow a power law when the time is measured from the moment at which the shock front reaches the surface. The power index is found to be determined by the condition for the flow to extend through a critical point. The energy spectrum of the ejected matter as a result of the shock breakout is derived and its dependence on the strength of the explosion is also deduced. The results are compared with the self-similar solution for the same problem with non-relativistic treatment. 
\end{abstract}


\section{Introduction}

In a supernova (SN) explosion of a compact star, a shock wave is accelerated to a relativistic speed in the surface layer with a sharp density gradient and results in relativistic motion of the ejecta \citep{Colgate, Matzner, Tan}. This kind of SNe is observationally classified into type Ic (SNe Ic). From radio observations for one of SNe Ic (SN 1998bw), \citet{Kulkarni} inferred that the radio shock front originated from this SN was moving at relativistic speeds with the Lorentz factor between 1.6 and 2.
Since the discovery of a close association of a $\gamma$-ray burst GRB 980425 with this supernova \citep{Galama, Iwamoto}, there has been accumulated evidence that some of SNe Ic are related to a class of $\gamma $-ray bursts. \citet{Rees}  proposed a plausible model  (the fireball shock model) to describe $\gamma$-ray bursts as phenomena taking place at cosmological distances, in which relativistic motion with the Lorentz factor of $\sim 100$ is required to accommodate the observed variability on very short time scales to huge $\gamma$-ray intensities and the shock converts its kinetic energy of the baryons into electrons to emit nonthermal $\gamma$-ray photons. Other models involving no relativistic motion would lead to a conflict that the size of the source indicated by the observed short time scales   are compact enough for $\gamma$-ray photons to collide to yield e$^+$e$^-$ pairs and emit little $\gamma$-rays. This model predicted {\it afterglows} of $\gamma$-ray bursts \citep{Meszaros}. The first detection of the X-ray afterglow from GRB 970228 by the Beppo-Sax satellite \citep{Costa} and the subsequent identification of the host galaxy confirmed the $\gamma$-ray burst as an event at a cosmological distance \citep{van}. Thus it will be important to investigate under what conditions the matter ejected from a SN Ic (or a core collapse SN in general) give rise to this relativistic motion.

Another aspect of the relativistic SN ejecta can be seen in the production of light elements.
Recent observations for metal-poor stars have shown that the abundance of Be increases proportionally to the abundance of Fe, which indicates that primary processes predominantly produce Be. It has been shown  \citep{Fields, Nakamura} that the relativistic motion of ejecta from SNe Ic may contribute to the production of light elements Li, Be, and B through spallation reactions as a primary process because the surface layer of the ejecta is composed of C and O. Though these studies are based on spherically symmetric models for SNe Ic, it would be interesting to investigate the effects of an aspherical SN explosion on the acceleration of the ejecta and the production of the light elements. \citet{Tan} and \citet{Nakamura} pointed out that an aspherical explosion may enhance the acceleration compared to a spherical explosion with the same explosion energy. Thus an aspherical explosion might produce more light elements. Furthermore, the acceleration strengthened by an aspherical explosion should be realized by a collimation of the flow. The collimated flow or jet is required to interpret the time evolution of the afterglow of some $\gamma$-ray bursts \citep{Rhoads}. Hence the connection between SNe and $\gamma$-ray bursts needs to be explored in this respect as well. A linear stability analysis of the corresponding spherical or plane-parallel flow will be a good starting point to understand this possible collimation mechanism. 

To study the acceleration of matter before and after the passage of a shock wave across the surface of a stratified medium adjacent to a vacuum in general, we will present a self-similar solution for an ultra-relativistic shock wave propagating in a plane-parallel medium. This problem without taking into account relativistic effects was posed by  \citet{Gandel} and fully solved by \citet{Sakurai}. Their solutions inevitably involve an infinite velocity at the surface after the shock emergence. Therefore a self-consistent solution needs to take into account relativistic effects.
To obtain a relativistic solution, we follow the procedure of \citet{Sakurai} with a different choice of non-dimensional independent variables.  As was done in \citet{Bland} for an ultra-relativistic shock wave emanating from a strong point explosion, the Lorentz factor of the shock front is assumed to evolve with time $|t|^{-m/2}$. For our problem, the time $t$ is assumed to be measured from the moment when the shock front hits the surface and the parameter $m$ is to be determined by the condition that the flow extend through a critical point (see \S \ref{BSE}). A similar work was studied by \citet{Perna} to describe the propagation of a relativistic shock wave in an exponential atmosphere. In this situation, the flow after the shock emergence cannot be treated because the surface is located at infinity. Another work related to the present work was done by \citet{Johnson}. They derived a solution for free expansion into a vacuum with the initial condition that the fluid is at rest while we will be concerned with a solution for free expansion of the shocked fluid moving into a vacuum. \citet{Johnson} also derived a solution for the shock propagation into a medium of decreasing density which starts from two adjacent fluids at rest with different pressures. They have treated the same problem as in the present paper except for the initial conditions. Thus the solutions will be compared in some detail.

  Since the assumed density distribution is realized when a polytropic gas is in hydrostatic equilibrium with a constant gravitational acceleration, there will be many situations to which this solution can apply in addition to the surface of a star. The analytical description of the flow presented in this paper will enable us to easily investigate the linear stability of the flow. Therefore this study is essential to investigating whether a disturbed shock front will enhance the acceleration of matter and lead to a collimated flow. Furthermore, this study will give a possible physical mechanism of the energy injection to the fireball model. 

Next section outlines the problem discussed in this paper. The flow before shock emergence is formulated and  its self-similar solution is derived in \S \ref{BSE}. The subsequent self-similar flow after shock emergence is described in \S \ref{ASE}. The analytical solution for the flow at extremely large $t$ is also presented. Based on these solutions, \S \ref{energyspectrum} discusses the energy spectrum of the outermost layer of the ejected matter as a result of the shock breakout into a vacuum. \S \ref{summary} concludes the paper. The Appendix gives the analytical description for the self-similar flow before shock emergence. 

\section{Outline of the problem}
A self-similar evolution of the flow associated with the ultra-relativistic shock wave in a plane parallel stratified medium and the subsequent flow after the shock emergence is investigated in the following context:
\begin{description}
\item[(a)] The initial density distribution $\rho_{0}(x)$ is given by a power law as 
\begin{equation}\label{density}
\rho_{0}(x) =
\begin{cases}
bx^{\delta}& \text{for $x\geq 0$},\\
0& \text{for $x<0$},
\end{cases}
\end{equation} 
where $x$ is the distance measured from the surface. Two positive constants $b$ and $\delta$ have been introduced. Note, however,  that this distribution is a good approximation only near the surface of a star because the  gravitational acceleration needs to be assumed constant and the sphericity is ignored.
We will consider two specific cases in some detail: a convective envelope  expressed with $\delta=1.5$ and radiative one with $\delta=3$, though our solutions presented in this paper hold for any positive $\delta$. 
The exterior of the star denoted by negative $x$ is assumed to be vacuum before the shock emergence. 
\item[(b)] An ultra-relativistic shock wave propagates toward the surface in a static medium with the density distribution given above.  The effect of gravity on the shock wave is assumed to be negligible. The pressure in front of the shock wave is neglected.
The time $t$ is measured from the instant of shock arrival at the surface. Thus this stage before the shock emergence is denoted by negative $t$. 
\item[(c)] After the shock emergence ($t>0$), the fluid expands into the vacuum. The initial conditions for the flow are given by the solution in the stage (b) when the shock hits the surface ($t\rightarrow -0$). The analytical expression for the asymptotic solution with $t\to\infty$ will be shown.

\end{description}

\section{Before Shock Emergence}
\label{BSE}
\subsection{Formulation}
In this section, we derive equations governing the propagation of  an ultra-relativistic shock wave and the associated self-similar flow from basic equations of relativistic hydrodynamics and discuss the boundary conditions at  the shock front. 
Next, we define the similarity variable and convert the 
partial differential equations into a set of ordinary differential equations.
The energy and momentum conservation laws in a fixed frame ($x,t$) can be expressed as 
  \begin{equation}  \label{em1} \frac{ \partial}{ \partial t}( \gamma ^2(e+ \beta ^2p))
        + \frac{ \partial}{ \partial x}( \gamma^2  \beta(e+p))=0    ,\end{equation}
  \begin{equation}   \frac{ \partial}{ \partial t}( \gamma ^2  \beta (e+p))
        +  \frac{ \partial}{ \partial x}( \gamma ^2  \beta^2 (e+p))+ \frac{ \partial p}{ \partial x}=0  ,
  \end{equation}
in a plane parallel atmosphere. Here $ \gamma $ denotes the Lorentz factor, $ \beta $ the velocity divided by the speed of light $c$ 
(we set $c=1$ in the following), and $p$ denotes the pressure. The energy density $e$ per unit volume includes the thermal energy and the rest mass energy.
 The continuity equation is written in terms of the number density $n^{ \prime}$ in the fixed frame as
 \begin{equation}  \label{cont}
 \frac{ \partial n^{ \prime}}{ \partial t} + \frac{ \partial}{ \partial x}( \beta n^{ \prime}) =0. \end{equation}
This density is defined by $n^\prime=n\gamma$ with the number density $n$ in the co-moving frame.
Equations (\ref{em1})--(\ref{cont}) are to be solved with an equation of state.
 Since we are concerned with the propagation of a shock wave in the relativistic limit, it will be appropriate to use the ultra-relativistic equation of state
 \begin{equation} \label{eos} e=3p .\end{equation}  
Using the relation between the partial time derivative $\partial /\partial t$ at a fixed point and the total derivative $d/dt$: $d/dt=\partial /\partial t+\beta \partial /\partial x$, together with this equation of state, 
equations (\ref{em1})--(\ref{cont}) are combined to yield 
 \begin{equation}  \label{dpdt} \frac{d}{dt}(p \gamma^4)= \gamma^2 \frac{ \partial p}{ \partial t} ,\end{equation}
 \begin{equation}   \frac{d}{dt} \ln (p^3  \gamma^4)=-4 \frac{ \partial  \beta}{ \partial x} ,\end{equation}
 \begin{equation}  \label{adiabat} \frac{d}{dt}(pn^{- \frac{4}{3}})=0 .\end{equation}

Then the non-dimensional pressure $f$, Lorentz factor $g$, and density $h$ are introduced as functions of the similarity variable $\chi$. They are defined to become unity at the shock front satisfying the ultra-relativistic Rankine-Hugoniot relations \citep[see e.g., ][]{Bland} as follows,
 \begin{equation} \label{rhp} p =  \frac{2}{3}bX_s^{ \delta}  \Gamma^2 f( \chi) ,\end{equation}
 \begin{equation}  \label{rhg} \gamma^2 =  \frac{1}{2} \Gamma^2 g( \chi) ,\end{equation}
 \begin{equation} \label{rhn} n^\prime = 2\frac{b}{\mu} X_s^{ \delta} \Gamma^2h( \chi) ,\end{equation}
where $X_s$ and $\Gamma$ are the position and  
the Lorentz factor of the shock front and 
$\mu$ is the mean mass of the constituent particles. As was mentioned in \S 1, the Lorentz factor of the shock front $\Gamma$ is assumed to evolve with the time $t$ as
\begin{equation}  \label{Lorentz} \Gamma ^2 = A(-t)^{-m} ,\end{equation}
where $A$ is a constant. Following \citet{Bland}, the similarity variable $\chi$ is defined as
 \begin{equation}  \label{chi}  \chi = \{ 1+2(m+1) \Gamma ^2\}(1+ \frac{x}{t}). \end{equation}
Here the power index $m$ must be greater than $-1$ so that the shock speed not exceed the speed of light. 
Note that when the shock reaches the surface, i.e. $t \to -0$, the Lorentz factor $\Gamma$ 
diverges, i.e., the shock speed approaches to the speed of light.
The value of $m$ can not be determined from the initial conditions, but from the condition that the solution extend across a critical point to infinity. 
The boundary conditions at the shock front are expressed as
 \begin{equation} \label{boundary} f(1)=g(1)=h(1)=1.\end{equation}

 The coordinate system can be transformed from $(x,\,t)$  to $(\Gamma,\,\chi)$ by the following relations
 \begin{equation} t \frac{ \partial}{ \partial t}=  - \frac{m \Gamma}{2} \frac{ \partial}{ \partial  \Gamma} 
                                                         +(m+1)(2 \Gamma^2 - \chi) \frac{ \partial}{ \partial  \chi},\end{equation}
 \begin{equation} t \frac{ \partial}{ \partial x}=2(m+1) \Gamma^2  \frac{ \partial}{ \partial  \chi}  ,\end{equation}
 \begin{equation} t \frac{d}{dt}=  - \frac{m \Gamma}{2} \frac{ \partial}{ \partial  \Gamma} 
                                   + (m+1)( \frac{2}{g}- \chi) \frac{ \partial}{ \partial  \chi} .  \end{equation} 
These relations hold to the accuracy of the order of $ \Gamma^{-2}$. Using these relations and substituting (\ref{rhp})-(\ref{rhn}) into equations (\ref{dpdt})-(\ref{adiabat}) yields the following ordinary differential equations,                                   
 \begin{equation}  \label{dfdx} \frac{1}{g} \frac{d \ln f}{d \chi} =  \frac{( \delta -m)g \chi+8m-4 \delta}{(m+1)(g^2 \chi^2-8g \chi+4)},\end{equation}
 \begin{equation}  \label{dgdx} \frac{1}{g} \frac{d \ln g}{d \chi} =  \frac{-mg \chi+7m-3 \delta}{(m+1)(g^2 \chi^2-8g \chi +4)} ,\end{equation}
 \begin{equation} \label{dhdx} \frac{1}{g} \frac{d \ln h}{d \chi} =  \frac{(m- \delta)g^2 \chi^2+(8\delta -10m)g \chi +18m-10 \delta} {(m+1)(2-g \chi)(g^2 \chi^2-8g \chi+4)}  .\end{equation}
 These equations have been expanded in powers of $\Gamma^{-2}$ and $\gamma^{-2}$ and truncated at the first contributions.  Given values of $ \delta$ and $m$, these equations are integrated from $\chi=1$ to $\chi=-\infty$ with the boundary conditions (\ref{boundary}).  However, it is found that the derivatives diverge at a finite $\chi$ except for a certain value of $m$. 
To avoid this, the denominators and numerators on the right hand sides of these equations  must vanish simultaneously, which occurs when
 \begin{equation} \label{eigenvalue} m =  \left(\mp 2\sqrt{3}-3\right)\delta.\end{equation}
The shock being accelerated toward the surface indicates a positive $m\left(=\left(2\sqrt{3}-3\right) \delta\right)$. If the negative value were chosen, the shock Lorentz factor would decrease to zero (less than 1) as the shock approaches the surface. This is not physical and the approximation of the relativistic limit would not hold. Hence, equation (\ref{dgdx}) is reduced to a simpler form,
\begin{equation} \frac{1}{g}\frac{d \ln g}{d\chi} = -\frac{\sqrt{3}\delta}{\left\{2+\sqrt{3}(\delta+1)\right\}\left(g\chi -4-2\sqrt3\right)} .\end{equation}
The other two equations are also simplified to equations (\ref{asdfdx}) and (\ref{asdhdx}) in the Appendix. It should be noted here that \citet{Johnson} have also presented an expression for the shock propagation in equation (43) of their paper, which indicates 
\begin{equation}
\Gamma\propto n^{-\frac{\sqrt3}{2(2+\sqrt3)}}.
\end{equation}
This relation is derived from an invariant quantity carried by forward characteristics and should hold in general situations.
Since the density $n$ at the shock front evolves with time $t$ as $n\propto (-t)^{-\delta}$, this expression is precisely equivalent to the shock propagation indicated by the value of $m$ in equation (\ref{eigenvalue}). 

\begin{figure}
\epsscale{.80}
\plotone{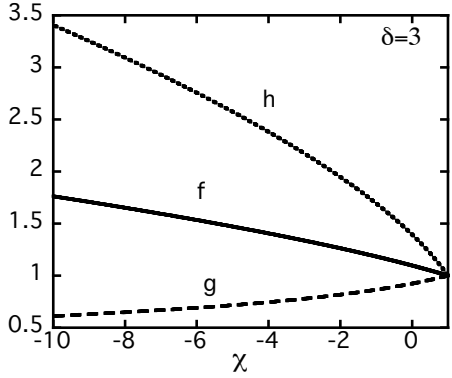}
\caption{The solution of equations (\ref{dfdx})-(\ref{dhdx}) for $\delta =3$ which corresponds to $m = 6\sqrt3-9\sim 1.392$ from equation (\ref{eigenvalue}). 
The integral curve $g$ passes through the singular 
point which is represented by $g\chi =4-2\sqrt3 $ without any special behavior.}
\label{bse3}
\end{figure}
\begin{figure}
\epsscale{.80}
\plotone{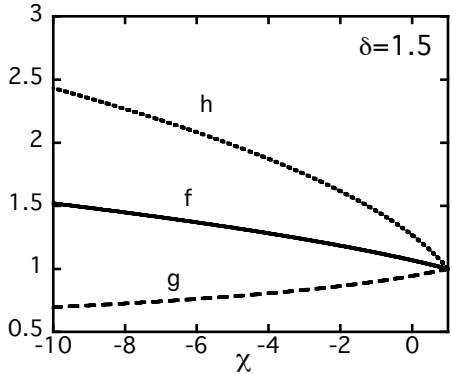}
\caption{The same as Figure \ref{bse3} but for $\delta =1.5$ which corresponds to $m =3\sqrt3-4.5\sim 0.696$. }
\end{figure}

\subsection{Solutions} \label{bsolu}
Though the analytical form obtained by integrations of (\ref{dfdx})--(\ref{dhdx}) is available and presented in the Appendix, the dependent variables $f$, $g$, and $h$ can be expressed in terms of $g\chi$ rather than the independent variable $\chi$. Thus we numerically integrate equations (\ref{dfdx})--(\ref{dhdx}) to obtain the distributions of the pressure, density, and Lorentz factor as functions of $\chi$. The results are shown in Figure 1 for $\delta =3$ and Figure 2 for $\delta =1.5$. This expression is more convenient for investigating the distribution and time evolution of the flow.

Here we will make a comparison of this solution with that presented by \citet{Johnson}. They have derived a relation between the Lorentz factors $\gamma$ and the positions $x$ of a fluid element at two different moments after passing through the ultra-relativistic shock front propagating in the medium with the density distribution given by equation (\ref{density}),
\begin{equation} \label{jmrelation} \frac{x}{x_0} = \frac{\left\{3+2\sqrt{3}\left(1+\delta\right)\right\}\left(\frac{\gamma_0}{\gamma}\right)^\frac{2(\sqrt{3}+2)}{\sqrt{3}\delta}-\left(\frac{\gamma}{\gamma_0}\right)^{(1+\sqrt{3})}}{2\left\{1+\sqrt{3}\left(1+\delta\right)\right\}}.  \end{equation}
Here $\gamma_0$ and $x_0$ denote the values at the moment when the fluid element is hit by the shock wave. Note that the original form in \citet{Johnson} is slightly different from the above expression because our coordinate $x$ is  different from theirs. We have confirmed that this relation can be deduced from our solution by the following procedure. In the relativistic limit, the distance $x$ from the surface is simply proportional to time $t$. Thus we obtain
\begin{equation} \label{position} \frac{x}{x_0}=\frac{t}{t_0}=\left(\frac{\Gamma_0}{\Gamma}\right)^\frac{2}{m}=\left(g\frac{\gamma_0^2}{\gamma^2}\right)^\frac{(\sqrt{3}+2)}{\sqrt{3}\delta}. \end{equation}
Here the time $t$ has been converted into the shock Lorentz factor $\Gamma$ using equation (\ref{Lorentz}).  Then the adiabatic evolution of the fluid element after passing the shock front yields a relation between the non-dimensional functions and the Lorentz factors,
\begin{equation} \label{entropy} g^\frac{-2}{3\sqrt{3}}fh^\frac{-3}{4}=\left(\frac{\gamma_0}{\gamma}\right)^\frac{4(1+\sqrt{3})}{3\sqrt{3}}. \end{equation}
Eliminating $g$, $f$, $h$ from this equation and equations (\ref{truef})--(\ref{trueh}) in the Appendix, $g\chi$ can be expressed as a function of the ratio of the Lorentz factors $\gamma_0\big/\gamma$. Substituting this expression into $\xi=-g\chi$ in equation (\ref{trueg}), then $g$ is expressed as 
\begin{eqnarray} \label{ggamma} \ln g = &\frac{\sqrt{3}\delta}{\sqrt{3}+2}&\ln\frac{\left\{3+2\sqrt{3}(1+\delta)\right\}\Lambda-1}{2\left\{1+\sqrt{3}(1+\delta)\right\}\Lambda},  \nonumber \\ &\mbox{ where }&\Lambda = \left(\frac{\gamma_0}{\gamma}\right)^\frac{(1+\sqrt{3})\left\{1+\sqrt{3}(1+\delta)\right\}}{\sqrt{3}\delta}.
\end{eqnarray}
Eliminating $g$ from equations (\ref{position}) and (\ref{ggamma}) yields the relation (\ref{jmrelation}). 
Thus the assumption of self-similarity does not lose any generality. The above relation (\ref{jmrelation}) can yield the ratio of the Lorentz factors $\gamma_0/\gamma$ of a fluid element when it passes the original surface by setting $x=0$. At this moment, the shock is already emerging from the surface. On the other hand,  the self-similar solution presented in this section holds as long as the shock exists. To describe the flow after the shock disappears as it hits the surface, an extended solution is developed in the following sections.

\subsection{Limiting Behavior}
In order to extend the solution obtained in the previous section to describe the flow after the shock emergence, the limiting behavior of the fluid when $t \to -0$ needs to be described.
From equation (\ref{chi}), the position of the shock front is given by the formula
 \begin{equation} X_s=-t \left\{ 1- \frac{1}{2(m+1) \Gamma^2} \right\}.\end{equation} 
 Hence $X_s \to -t$ as $t \to -0$. Thus the behavior of $f$, $g$, and $h$ defined by  (\ref{rhp})--(\ref{rhn}) is determined by the condition that $p$,  $\gamma$, and $n^{ \prime}$ must not vanish or diverge at all places with finite $\chi$'s. To satisfy this condition, we need to eliminate their dependences on $t$ to yield
 \begin{equation} \label{fasympt} f_1( \chi)  \propto (- \chi)^ \frac{\delta-m}{m+1}\propto (- \chi)^{ \frac{ 2\delta}{2+\sqrt 3(1+\delta)}}   ,\end{equation}
 \begin{equation} g_1( \chi) \propto (- \chi)^ \frac{-m}{m+1}  \propto (- \chi)^{- \frac{\sqrt 3\delta}{2+\sqrt 3(1+\delta)}}  ,\end{equation}
 \begin{equation} \label{hasympt} h_1( \chi)   \propto (- \chi)^ \frac{\delta-m}{m+1}\propto (- \chi)^{ \frac{ 2\delta}{2+\sqrt 3(1+\delta)}}  , \end{equation}
where the subscript 1 indicates the limiting functions. 
In the Appendix, this argument is verified in the analytical form. 
As a result, their distributions at the instant of the shock emergence are given by
 \begin{equation} \label{plim} p_1(x) = \frac{2Ab}{3}\left[\frac{2A\left\{1+(2\sqrt3-3)\delta\right\}}{3+2\sqrt3(1+\delta)}\right]^{ \frac{ 2\delta}{2+\sqrt 3(1+\delta)}}x^{ \frac{ 2\delta}{2+\sqrt 3(1+\delta)}} ,\end{equation}
 \begin{equation}  \label{gamma1}\gamma_1^2(x) = \frac{A}{2}\left\{\frac{12+7\sqrt3+2\left(2\sqrt3+3\right)\delta}{2\left[2+\sqrt3(1+\delta)\right]A}\right\}^\frac{\sqrt 3\delta}{2+\sqrt 3(1+\delta)} x^{- \frac{\sqrt 3\delta}{2+\sqrt 3(1+\delta)}} , \end{equation}
 \begin{equation} \label{nlim} n_1^{ \prime}(x) = \frac{2Ab}{\mu}\left[\frac{2A\left\{1+(2\sqrt3-3)\delta\right\}}{\left\{3+2\sqrt3(1+\delta)\right\}^\frac{\sqrt3(1+\delta)}{2\left\{1+\sqrt3(1+\delta)\right\}}}\right]^\frac{ 2\delta}{2+\sqrt 3(1+\delta)} x^{ \frac{ 2\delta}{2+\sqrt 3(1+\delta)}} .\end{equation}
Here the subscript 1 indicates distributions at $t=0$. The distribution of the density has a clearly different power law index from that in the initial condition. This is a purely relativistic effect, since it does not happen in non-relativistic case  \citep{Sakurai}, in which the strong shock always compresses the fluid by a constant factor.
  
\section{After Shock Emergence}
\label{ASE}
\subsection{Formulation} 

Since no new scale is introduced after the shock emergence, the fluid keeps its self-similarity \citep[in Newtonian case, see][]{Sakurai}. Lagrangian coordinates are convenient for describing the motion of the expanding gas into a vacuum starting from given initial distributions \citep{Sakurai}.  Thus we will use a Lagrangian coordinate $a$ defined as it coincides with the Eulerian coordinate $x$ at $t=0$ in the previous section. The relation between $x$ and $a$ is given by
 \begin{equation}  \label{dxda} \frac{ \partial x}{ \partial a} =  \frac{n_1^{\prime}}{n^{\prime}} .\end{equation}
Now we seek a solution in the form of
 \begin{equation} \label{pafter} p = p_1(a)F( \eta),\end{equation}
 \begin{equation}  \gamma^2 =  \gamma_1^2(a)G( \eta) ,\end{equation}
 \begin{equation} \label{nafter} n = n_1(a)H( \eta),\end{equation}
 \begin{equation} x = aR( \eta)  ,\end{equation}
where the similarity variable $ \eta$ is defined as
 \begin{equation}  \label{afsimv} \eta =  \frac{t}{ \gamma_1^2 a}  \propto ta^{-\frac{1}{m+1}}\propto ta^{ -\frac{1}{1+(2\sqrt 3-3)\delta}}. \end{equation}
The initial conditions are 
 \begin{equation} \label{afinit} F(0) = G(0) = H(0) = R(0)=1.\end{equation} 
Using the following transformations from $(t, a)$ to $(\eta, a)$ (here, we express the total derivative $d/dt$ 
as the partial derivative $ \partial / \partial t$ in order to emphasize that it is taken with $a =$const.).
 \begin{equation}  t\frac{ \partial}{ \partial t} \to  \eta  \frac{ \partial}{ \partial  \eta},\end{equation}
 \begin{equation} a \frac{ \partial}{ \partial a} \to - \frac{\eta}{m+1}  \frac{ \partial}{ \partial  \eta} +a\frac{\partial }{\partial a}, \end{equation}
and (\ref{dxda}), equations (\ref{dpdt})-(\ref{adiabat}) are converted to the following equations in the relativistic limit,
 \begin{equation}  \label{dgdeta} \frac{G^{ \prime}}{G} =  -\frac{3F^{\prime}}{2F}-\frac{H}{(m+1)\sqrt{G}}\left(\frac{\eta G^{\prime}}{G}+m\right),\end{equation}
\begin{equation}  \label{dfdeta} \frac{F^{ \prime}}{F} =  \frac{( \delta -m)H\sqrt{G}-2(m+1)G^{ \prime}}
      {(m+1)G+ \eta H\sqrt{G}}  ,\end{equation}
 \begin{equation} \label{dhdeta} \frac{H^{ \prime}}{H} =  \frac{3F^{ \prime}}{4F}  ,\end{equation}
 \begin{equation} \label{drdeta}  \frac{R^{ \prime}}{R} =  \frac{m+1}{ \eta}\left(1-\frac{1}{RH\sqrt{G}} \right),\end{equation}
 where the prime denotes the derivative with respect to $ \eta$ and $m=(2\sqrt 3-3)\delta$ as in (\ref{eigenvalue}).

\begin{figure}
\epsscale{.80}
\plotone{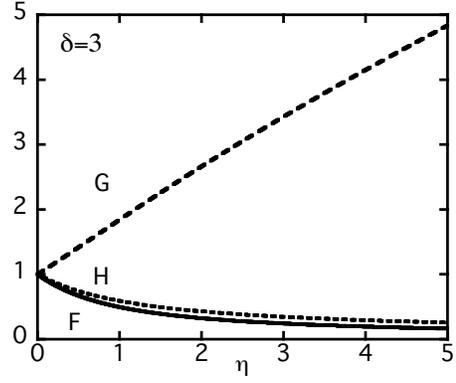}
\caption{The solution of equations (\ref{dgdeta})-(\ref{dhdeta}) for $\delta =3$. }
\label{ase3}
\end{figure}

\begin{figure}
\epsscale{.80}
\plotone{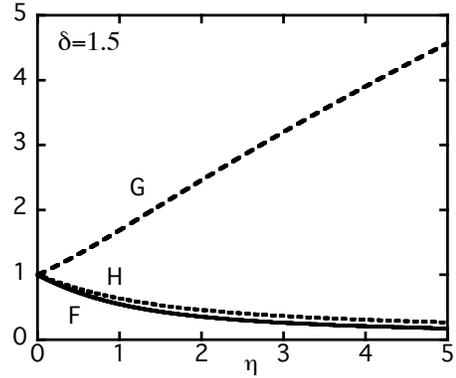}
\caption{The same as Figure \ref{ase3} but for $\delta =1.5$. }
\end{figure}

\subsection{Solutions}\label{fsolu}
The ordinary differential equations (\ref{dgdeta})--(\ref{drdeta}) are numerically integrated with the initial conditions (\ref{afinit}).  Unlike the flow before shock emergence, the solution does not have a critical point. The results  for $\delta = 3$ and 1.5 are shown in Figures 3 and 4.

There are two invariants in this solution. One is from the adiabatic condition (eq. (\ref{dhdeta})), 
\begin{equation} \label{inv1} F(\eta)H(\eta)^{-4/3}=1, \end{equation}
and the other is 
\begin{equation} \label{inv2} G(\eta)F(\eta)^{\sqrt3/2}=1,\end{equation}
obtained by eliminating $H(\eta)$ from equations (\ref{dgdeta}) and (\ref{dfdeta}). This is the Riemann invariant whose precise form in  the relativistic limit is given in \citet{Johnson}. 
From equations (\ref{dgdeta})-(\ref{drdeta}), we can derive the asymptotic solution in the limit of $\eta \to \infty$ undergoing power  law evolution :
\begin{equation} \label{aform} F(\eta)=F_1\eta^\phi,\, G(\eta)=G_1\eta^\psi, \,H(\eta)=H_1\eta^\theta .\end{equation}
Substituting  them into  equations  (\ref{dgdeta})-(\ref{dhdeta}), it follows that the power law indices must satisfy the following relations 
\begin{equation} \psi = -\frac{3}{2}\phi-\frac{d}{m+1}(\psi+m) ,\end{equation}
\begin{equation} \phi = \frac{(\delta-m)d-2(m+1)\psi}{m+1+d} ,\end{equation}
\begin{equation} \theta = \frac{3}{4}\phi ,\end{equation}
\begin{equation} \frac{\psi}{2} - \theta =1 ,\end{equation}
where $d=H_1/\sqrt{G_1}$. These equations have two solutions one of which is independent of $\delta$ or $m$ except for $d$. The other solution gives a negative $\psi$ which is unphysical. Thus the time scale of the asymptotic behavior does not depend upon the initial conditions because the leading edge of the fluid expands as fast as the speed of light.
As a result, we obtain 
\begin{equation*} \label{afindex} \phi=2(\frac{1}{\sqrt3}-1), \,\psi=\sqrt3-1, \,\theta=\frac{\sqrt3-3}{2}, 
\end{equation*}
\begin{equation}
\mbox{and }   d=\frac{2(2-\sqrt3)\{\sqrt 3(\delta+1)+2\}}{1+\sqrt3(\delta+1)} .\end{equation}
Using these power law indices, the non-dimensional position $R$ is found to evolve as $\eta^{m+1}$ from equation (\ref{drdeta}). We have confirmed that these are exactly matched with the results of the numerical integration. 
Furthermore, we can express the coefficients $F_1,G_1$ and $H_1$ in terms of $\delta$ using equations (\ref{inv1}) and (\ref{inv2}) together with the expression for $d$ in (\ref{afindex}) as 
\begin{equation} \label{afcoeff} F_1=d^{2\left(1-1/\sqrt3\right)}, \,G_1=d^{1-\sqrt3}, \,
       H_1=d^{\frac{8}{3}\left(1-{1/\sqrt3}\right)}.\end{equation}
These values also agree with our numerical results. 
All the coefficients depend on $\delta$ and remember the initial density distribution.

Our results are similar to those of \citet{Johnson} in some sense, though their initial conditions are different from ours. For example, it follows from our results that $\gamma_f^2 = \gamma_1^2 (p_1/p_f)^{\sqrt3/2}$, where the subscript $f$ denotes the final value. On the other hand, \citet{Johnson} show $4\gamma_f^2 = (p_1/p_f)^{\sqrt3/2}$. Their fluid is initially static and uniform while ours is already not uniform and expanding at this initial moment, though both treat a fluid expanding into a vacuum.  Due to their initial conditions, their problem inevitably involves non-relativistic flow in which the Riemann invariant obtained above in the relativistic limit is not appropriate. This causes the difference in the factor of 4 in front of $\gamma^2_f$.  Our approach can be applied to any small value of $\delta$ as long as it is positive, because the shock front is eventually accelerated to ultra-relativistic speeds.  It is, however, not appropriate for $\delta =0$,  a special case of which is treated by \citet{Johnson}. 

\begin{deluxetable}{cccccc}
\tabletypesize{\scriptsize}
\tablecaption{Power-law indices for $p$ and $n$.}
\tablewidth{0pt}
\tablehead{
 & &\multicolumn{2}{c}{$\partial\ln p/\partial\ln a$} & \multicolumn{2}{c}{$\partial\ln n/\partial\ln a$} \\
\raisebox{1.5ex}[0pt]{$\delta$}& \raisebox{1.5ex}[0pt]{$t$} & \colhead{relativistic} & \colhead{non-relativistic} & \colhead{relativistic} & \colhead{non-relativistic} 
}
\startdata
  & 0       & 0.67 & 1.89 &1.09 & 3.00\\
\raisebox{1.5ex}[0pt]{3}  & +$\infty$ & 1.03 & 3.96 &1.23 & 4.56 \\
& 0       & 0.47 & 0.84 & 0.85 & 1.50\\ 
\raisebox{1.5ex}[0pt]{1.5}  & +$\infty$ & 0.97 & 2.54 & 1.05 & 2.79 \\
\enddata
\end{deluxetable}

\subsection{Asymptotic Distribution}
The  distributions of physical quantities in the fluid in the limit of $\eta\to \infty$ are discussed in this section. The distributions of our self-similar solution converge to power law distributions in this limit, which is referred to {\it the asymptotic distribution} here.  This distribution is still intermediate and eventually changes to non-self-similar distribution when the matter cools down and no longer behaves as ultra-relativistic gas.  Of course, a real star is spherical thus sphericity should be taken into account 
when we consider the motion far from the stellar surface. The effects of sphericity will be incorporated by expanding the flow variables in the power series of the ratio of the distance from the surface to the stellar radius  \citep{Kazhdan} or by introducing an empirical formula obtained from comparisons with numerical calculations  for spherically symmetric explosions \citep{Matzner}.

When $t\to \infty$, i.e. $\eta \to \infty$, the asymptotic distributions of $F, G$ and $H$ have been already 
obtained in (\ref{afindex}) and (\ref{afcoeff}). Thus  the pressure, the Lorentz factor, and the density  have power  law distributions with respect to $a$ as
\begin{equation}  \label{pasympt} p\propto a^{\frac{2(\sqrt3\delta+1+\sqrt3)}{2\sqrt3+3(\delta+1)}}, \end{equation}
\begin{equation} \gamma^2 \propto a^{-\frac{1+\sqrt3(\delta+1)}{2+\sqrt3(\delta+1)}}, \end{equation}
\begin{equation} \label{nasympt} n \propto a^{\frac{(4+\sqrt3)\delta+\sqrt3(1+\sqrt3)}
                                                         {2\left\{2+\sqrt3(\delta+1)\right\}}}. \end{equation}
Numerically, the pressure and the density have distributions with respect to $a$ as
\begin{equation} p \propto a^{1.03} ,  n \propto a^{1.23}   (\delta = 3),\end{equation}
\begin{equation} p \propto a^{0.97} ,  n \propto a^{1.05}   (\delta = 1.5),\end{equation}
for specific values of $\delta$.
Comparing these results with those at the instant of shock emergence, 
there is only a small difference in the power law indices in the density distributions as compared with the difference in the pressure distribution (see Table 1). 

In non-relativistic case, however, the distributions significantly change after the shock emergence. 
According to Sakurai's solutions for a gas with the adiabatic index 4/3 \citep{Sakurai}, 
the distributions of pressure and density in the limit of $t\to \infty$ are 
\begin{equation} p \propto a^{3.96} ,  n \propto a^{4.56} (\delta =3),\end{equation}
\begin{equation} p \propto a^{2.54} ,  n \propto a^{2.79} (\delta = 1.5). \end{equation}
These power law indices  are quite different from those at the instant of shock emergence ($t=0$), 
details of the indices are shown in Table 1. 

The asymptotic behavior of $G$ suggests that the Lorentz factor continues to increase without bound as long as the ultra-relativistic equation of state (\ref{eos}) holds. The adiabatic expansion will eventually invalidate this equation of state and cease a further acceleration. This effect will be taken into account to obtain the energy spectrum of the ejecta in the next section.

\section{Energy Spectrum}
\label{energyspectrum}
From the self-similar solution discussed in the preceding sections, the energy spectrum in the ejected matter near the leading edge and its dependence on the initial strength of the explosion will be derived in this section. The initial strength of the explosion is deduced from the coefficient $A$ in equation (\ref{Lorentz}). It is expected that $A$ is proportional to the square of the explosion energy per unit mass, though both of the total energy and ejected mass involved in this self-similar solution are infinite. 

The energy per unit mass excluding the rest mass energy is given by
\begin{equation} \label{especific}\epsilon = \frac{\tau}{\rho \gamma} ,\end{equation}
where
\begin{equation} \tau = (\rho +4p)\gamma^2 - p - \rho \gamma.\end{equation}
Here we have used the ultra-relativistic equation of state (\ref{eos}). 

The energy spectrum is defined as the total mass with the energy per unit mass greater than $\epsilon$, 
\begin{eqnarray} \label{es} M(>\epsilon) &=& \int_{0}^{a(\epsilon)}\mu n_1^\prime(x)\,dx \propto
\left(a(\epsilon)A\right)^\frac{\delta+1}{m+1} \\ \nonumber &\propto&   
(a(\epsilon)A)^\frac{(2+\sqrt3)(\delta+1)}{2+\sqrt3(1+\delta)},\end{eqnarray}
where $a(\epsilon)$ denotes the Lagrangian coordinate of the fluid element with the energy per unit mass of $\epsilon$.  Substituting (\ref{plim})-(\ref{nlim}) (for $t=0$) or (\ref{pasympt})-(\ref{nasympt}) (for $t\to\infty$) into (\ref{especific}) and taking the dominant term at a specific time, we can evaluate the relation between $a$ and $\epsilon$. 
The energy distribution will be derived for limiting cases of $t=0$ and $t=\infty$ here
and also compared with Sakurai's non-relativistic cases for $\delta = 1.5$ and 3. 

At the instant of shock emergence, i.e., $t = 0$ , the specific energy $\epsilon$ in (\ref{especific}) becomes dominated by the term $p\gamma/\rho$ near the edge of the ejected matter.
Thus using the expressions for $p$, $\gamma$, and $\rho$ given by (\ref{plim})-(\ref{nlim}), we obtain 
\begin{equation} a(\epsilon) \propto A^\frac{1}{m}\epsilon^{-\frac{(m+1)}{m}} \propto A^{\frac{2+\sqrt3}{\sqrt3\delta}} \epsilon ^{-\frac{\left\{2+\sqrt3(1+\delta)\right\}}{\sqrt3\delta}} .\end{equation}
Then combining this result with equation (\ref{es}) yields 
\begin{equation} M(>\epsilon) \propto \left(\frac{A}{\epsilon}\right)^\frac{\delta+1}{m}\propto \left(\frac{A}{\epsilon}\right)^\frac{(\delta+1)(2+\sqrt3)}{\sqrt3\delta}.\end{equation}
For specific values of $\delta$, the spectrum becomes
\begin{equation} M(>\epsilon)\propto \begin{cases}
A^{2.87} \epsilon^{-2.87} & \text{for $\delta=3$,} \\
A^{3.59} \epsilon^{-3.59} & \text{for $\delta=1.5$.}
\end{cases}
\end{equation}
On the other hand, Sakurai's solution from non-relativistic treatments gives
\begin{equation} \label{sakuraisp} M(>\epsilon) \propto  \begin{cases}
A_{\rm nr}^{7.18} \epsilon^{-3.59} & \text{for $\delta = 3$,} \\ 
A_{\rm nr}^{8.73} \epsilon^{-4.37} & \text{for $\delta = 1.5$,}
\end{cases}
\end{equation}
where $A_{\rm nr}$ represents the strength of the initial explosion, 
defined as $V_s=A_{\rm nr}X_s^{-\lambda}$, where $V_s$ and $X_s$ denotes 
the velocity and position of the shock front, $\lambda$ is the eigen value  
determined by the method presented by \citet{Sakurai}. Thus this coefficient $A_{\rm nr}$ is expected to scale with the square root of the explosion energy per unit mass.

The distributions of $F$, $G$ and $H$ in the limit of $t\to \infty $
are given by (\ref{aform}),  (\ref{afindex}), and (\ref{afcoeff}). 
In this limiting situation, the internal energy has been completely converted to the kinetic energy. Thus $\epsilon\sim\gamma$.   Using these asymptotic behavior, the same procedure leads to 

\begin{equation} a(\epsilon) \propto A^\frac{\sqrt3-2}{1-\sqrt3-m}\epsilon^{\frac{2(m+1)}{1-\sqrt3-m}}\propto A^\frac{1}{1+\sqrt3(1+\delta)} \epsilon ^{-\frac{2\left\{2+\sqrt3(1+\delta)\right\}}{1+\sqrt3(1+\delta)}} ,\end{equation}
and
\begin{equation}  \label{esrelan}M(>\epsilon) \propto A^\frac{(2+\sqrt3)(\delta+1)}{1+\sqrt3(\delta+1)}\epsilon^{-\frac{2(2+\sqrt3)(\delta+1)}{1+\sqrt3(\delta+1)}}.\end{equation}
Numerically, 
\begin{equation} M(>\epsilon) \propto  \begin{cases}
A^{1.88} \epsilon ^{-3.77} & \text{for $\delta = 3$,} \\
A^{1.75} \epsilon ^{-3.50} & \text{for $\delta = 1.5$.}
\end{cases}
\end{equation}
The smaller power index of $\epsilon$ indicates that the acceleration is more effective in the inner part after the shock emergence. This phenomenon is not seen in non-relativistic treatments.
In Sakurai's non-relativistic solution, we obtain the energy spectrum with exactly the same dependence on $A_{\rm nr}$ and $\epsilon$ as at $t=0$. This solution shows that the energy in the outermost layer of the ejected matter is dominated by the kinetic energy at both $t=0$ and $t \to\infty$ and that the velocity profile in this region converges to one with the same power law index in both of these two limits. Therefore, it shows the final energy spectrum of the ejected matter with the same dependence on $A_{\rm nr}$ and $\epsilon$ as at $t=0$. Thus \citet{Fields2} used the relation (\ref{sakuraisp}) for $\delta = 3$ to derive the empirical formula for the energy spectrum of the ejected matter $M(>\epsilon) \propto \epsilon^{-3.6}\left(E_{\rm ex}/M_{\rm ej}\right)^{3.6}$, where the explosion energy per unit mass is expressed by the ratio of the explosion energy $E_{\rm ex}$ to the ejected mass $M_{\rm ej}$.
 
The above energy spectrum (\ref{esrelan}) as a result of the relativistic shock breakout is still intermediate and the final energy spectrum should be modified from it to that at the moments when the acceleration ceases in each fluid element. Assuming that the pressure suddenly drops to zero when $e/\rho$ reaches 1 \citep{Johnson}, the Lorentz factor $\gamma$ at this moment in each fluid element can be obtained as a function of $a$. Thus the final energy spectrum will change to 
 \begin{equation}\label{esrela}
M(>\epsilon) \propto A^\frac{(2+\sqrt3)(1+\delta)}{\sqrt3\delta}\epsilon^{-\frac{(3+\sqrt3)(\delta+1)}{3\delta}}.
\end{equation}
This leads to $M(>\epsilon) \propto \epsilon^{-2.10}\left(E_{\rm ex}/M_{\rm ej}\right)^{5.75}$ for $\delta=3$.  For $\delta =1.5$ the energy spectrum becomes $M(>\epsilon) \propto \epsilon^{-2.63}\left(E_{\rm ex}/M_{\rm ej}\right)^{7.18}$. On the other hand, the non-relativistic treatment yields the energy spectrum $M(>\epsilon) \propto \epsilon^{-4.4}\left(E_{\rm ex}/M_{\rm ej}\right)^{4.4}$  for $\delta=1.5$ by substituting $A_{\rm nr}\propto \sqrt{E_{\rm ex}/M_{\rm ej}}$ into the second equation of (\ref{sakuraisp}), which coincides with equation (24) of \citet{Matzner}. 

\section{Conclusions and discussion}
\label{summary}
We have presented a self-similar solution for a planar flow associated with the ultra-relativistic shock wave propagating in a power-law density distribution adjacent to a vacuum. The solution can describe the flow before and after the shock front reaches the surface. The solution at the shock emergence is explicitly given in the analytical form as the initial condition for the flow after the shock emergence. The asymptotic distributions of the flow as a result of the phenomenon are also presented in the analytical form. 

Thanks to the detailed comparison with the results of \citet{Johnson} presented in \S \ref{bsolu}, one can trace back each fluid element in the final state to the original position. Equation (\ref{jmrelation}) applied to the flow at $t=0$ yields the ratio of the Lagrangian coordinate $a$ to the original position $x_0$ as 
\begin{equation}
\frac{a}{x_0}=\frac{\left\{3+2\sqrt3(1+\delta)\right\}\left(\frac{\gamma_0}{\gamma_1}\right)^\frac{2(\sqrt3+2)}{\sqrt3\delta}-\left(\frac{\gamma_1}{\gamma_0}\right)^{(1+\sqrt3)}}{2\left\{1+\sqrt3(1+\delta)\right\}}
\end{equation}
These two coordinates are related to each other by the expression for $\gamma_1$ obtained in equation (\ref{gamma1}) in terms of the Lagrangian coordinate $a$ and $\gamma_0^2=Ax_0^{(3-2\sqrt3)\delta}/2$ as indicated by equation (\ref{rhg}). Therefore the flow in the final state described in terms of $a$ in \S \ref{fsolu} can be expressed as functions of $x_0$. 

It would be possible to formulate the flow before shock emergence with a Lagrangian coordinate as was shown by \citet{Matzner} for non-relativistic flow. However, it might not be possible to derive the eigen value of $m$ analytically from this formalism because the resultant differential equation corresponding to equation (\ref{dgdx}) would inevitably involve the self-similar variable for density. Furthermore, it is not obvious how to choose an appropriate self-similar coordinate. 

The energy spectrum in the ejected matter is deduced from the solution in terms of the mass $M(>\epsilon)$ of the ejected matter with the Lorentz factor greater than a certain value. The power law index of this mass with respect to $A$ greater than $1/2$ indicates that the enhancement of the initial explosion energy will accelerate the matter near the leading edge more effectively than that in the inner part of the ejected matter because this relation holds only in the outermost layers of the ejecta. Otherwise, the total energy would not be conserved. It is also expected from this power law index that if the explosion energy is not distributed uniformly in the lateral direction, the non-uniformity will be enhanced near the leading edge of the ejected matter and result in jets. This solution will be useful to test relativistic hydrodynamical codes whether they are appropriate to treat the passage of a shock wave across the surface of the matter to a vacuum.

 \acknowledgments
This work has been partially supported by the grant in aid (16540213) of the Ministry of Education, Science, Culture, and Sports in Japan.
   
\appendix

\section{Analytical Solutions Before Shock Emergence}

Differential equations (\ref{dfdx})-(\ref{dhdx}) can be analytically integrated after some manipulations.
Using a new independent variable defined as  $\xi = -g\chi$, equations (\ref{dfdx})-(\ref{dhdx}) are transformed into differential equations with respect to $\xi$,
\begin{equation} \label{adfdx}\frac{d\ln f}{d\xi} = -\frac{(m-\delta)\xi+8m-4\delta}{\xi^2+(3\delta +m+8)\xi+4(m+1)}, \end{equation}
\begin{equation} \frac{d\ln g}{d\xi}= -\frac{m\xi+7m-3\delta}{\xi^2+(3\delta +m+8)\xi+4(m+1)} ,\end{equation}
\begin{equation} \label{adhdx}\frac{d\ln h}{d\xi}= -\frac{(m-\delta)\xi^2+(10m-8\delta)\xi+18m-10\delta}
                                                                  {(\xi+2)\{\xi^2+(3\delta+m+8)\xi+4(m+1)\}}. \end{equation}
These equations can be integrated to yield 
\begin{equation} \label{af} f = \left|\frac{\xi-\alpha_1}{1+\alpha_1}\right|^{\frac{(\delta-m)(1+C_1)}{2}}
            \left|\frac{\xi-\alpha_2}{1+\alpha_2}\right|^{\frac{(\delta-m)(1-C_1)}{2}} ,\end{equation}
\begin{equation} g= \left|\frac{\xi-\alpha_1}{1+\alpha_1}\right|^{\frac{-m(1+C_2)}{2}}
            \left|\frac{\xi-\alpha_2}{1+\alpha_2}\right|^{\frac{-m(1-C_2)}{2}} ,\end{equation}
\begin{equation} \label{ah} h= \left|\frac{\xi-\alpha_1}{1+\alpha_1}\right|^{C_3(\delta-m)}           
            \left|\frac{\xi-\alpha_2}{1+\alpha_2}\right|^{C_4(\delta-m)}|\xi+2|^{C_5(\delta-m)} ,\end{equation}            
where constants have been introduced and are defined as
\begin{equation} \alpha_{1,2} = \frac{-(3\delta +m+8)\pm \sqrt{(3\delta +m+8)^2-16(m+1)}}{2} ,\end{equation}
\begin{equation} C_1 = \frac{-(3\delta +m+8)+8(2m-1)/(m-\delta)}{\alpha_1-\alpha_2} ,\end{equation}
\begin{equation} C_2 = \frac{-3\delta -m+6(1-\delta/m)}{\alpha_1-\alpha_2} ,\end{equation}
\begin{equation} C_3=\frac{(\alpha_1-2)(m^2+3\delta^2-4\delta m-25m+19\delta)+2(m+2)
                         (m^2-3\delta^2+2\delta m-2m)}
                         {2(m+2)(m-\delta)(\alpha_1-\alpha_2)} ,\end{equation}
\begin{equation} C_4 = \frac{-m^2+3\delta^2-2\delta m+2m}{(\alpha_1-2)(m-\delta)}-
                  C_3\frac{\alpha_2-2}{\alpha_1-2}, \end{equation}
\begin{equation} C_5=1-C_3-C_4 .\end{equation}
Here we have used the initial conditions (\ref{boundary}).
From (\ref{af})-(\ref{ah}), we can obtain the behaviors of $f$, $g$ and $h$ in the limit of $\xi \to +\infty$, 
i.e., $\chi \to -\infty$ where the constant terms $\alpha_1$ and $\alpha_2$ in these equations are
negligible. Thus we obtain 
\begin{equation} f \propto \xi^{\delta -m} , \,  g \propto \xi^{-m}  ,\,  h\propto \xi^{\delta-m} ,\end{equation}
which are rewritten in terms of $\chi$ as
\begin{equation} f \propto (-\chi)^{\frac{\delta-m}{m+1}} ,\,   g \propto (-\chi)^{-\frac{m}{m+1}}, \,
       h\propto (-\chi)^{\frac{\delta-m}{m+1}} .  \end{equation}
These results justify our anticipation for the limiting behavior of the solution (\ref{fasympt})-(\ref{hasympt}).

For the eigenvalue $m=\sqrt{3}\left(2-\sqrt{3}\right)\delta$ obtained in \S\ref{BSE}, equations (\ref{adfdx})--(\ref{adhdx}) are reduced to 
\begin{equation} \label{asdfdx}
\frac{d\ln f}{d\xi} = \frac{2\left(2-\sqrt{3}\right)\delta}{\xi+4+2\sqrt{3}\left(1+\delta\right)}, \end{equation}
\begin{equation} \frac{d\ln g}{d\xi} = \frac{\sqrt{3}\left(\sqrt{3}-2\right)\delta}{\xi+4+2\sqrt{3}\left(1+\delta\right)}, \end{equation}
\begin{equation} \label{asdhdx} \frac{d\ln h}{d\xi} = \frac{\left(2-\sqrt{3}\right)\delta}{1+\sqrt{3}\left(1+\delta\right)}\left\{\frac{2+\sqrt{3}}{\xi+2}+\frac{\sqrt{3}(1+2\delta)}{\xi+4+2\sqrt{3}(1+\delta)}\right\}. \end{equation}
These can be integrated to the analytical expression for the numerical solution presented in \S\ref{BSE}. By directly substituting the eigenvalue of $m$ into equations (\ref{af})--(\ref{ah}), we also obtain the analytical expressions,
\begin{equation}  \label{truef} \ln f = 2\left(2-\sqrt{3}\right)\delta\ln\left|\frac{\xi+4+2\sqrt{3}(1+\delta)}{3+2\sqrt{3}(1+\delta)}\right|, \end{equation} 
\begin{equation} \label{trueg} \ln g =\left(3-2\sqrt{3}\right)\delta\ln\left|\frac{\xi+4+2\sqrt{3}(1+\delta)}{3+2\sqrt{3}(1+\delta)}\right|, \end{equation} 
\begin{equation} \label{trueh} \ln h =\frac{\delta}{1+\sqrt{3}\left(1+\delta\right)}\left\{\ln\left|\xi+2\right|+\left(2\sqrt{3}-3\right)(1+2\delta)\ln\left|\frac{\xi+4+2\sqrt{3}(1+\delta)}{3+2\sqrt{3}(1+\delta)}\right|\right\}. \end{equation}


\begin{thebibliography}{}
\bibitem[Blandford  \& McKee(1976)]{Bland}Blandford, R.D. \& McKee, C. F. 1976,
    Phys. Fluids, 19, 1130
\bibitem[Colgate, McKee, \& Blevins(1972)]{Colgate} Colgate, 
S.~A., McKee, C.~R., \& Blevins, B.\ 1972, \apjl, 173, L87
\bibitem[Costa et al.(1997)]{Costa} Costa, E., et al.\ 1997, 
\nat, 387, 783
\bibitem[Fields, Daigne, Cass{\' e}, \& 
Vangioni-Flam(2002)]{Fields} Fields, B.~D., Daigne, F., 
Cass{\' e}, M., \& Vangioni-Flam, E.\ 2002, \apj, 581, 389
\bibitem[Galama et al.(1998)]{Galama} Galama, T.~J., et al.\ 
1998, \nat, 395, 670 
\bibitem[Gandel'man \& Frank-Kamenetskii(1956)]{Gandel} Gandel'man, G. M. \&
    Frank-Kamenetskii, D. A.  1956, Soviet Phys. Dokl., 1,223
\bibitem[Iwamoto et al.(1998)]{Iwamoto} Iwamoto, K., et al.\ 
1998, \nat, 395, 672 
\bibitem[Johnson \& McKee(1971)]{Johnson} Johnson, M.~H.~\& McKee, C.~F.\ 1971, \prd, 3, 858 
\bibitem[Kazhdan \& Murzina(1992)]{Kazhdan} Kazhdan, I.~M., \& 
Murzina, M.\ 1992, \apj, 400, 192 
\bibitem[Kulkarni et al.(1998)]{Kulkarni}
Kulkarni, S.~R.~et al.\ 1998, \nat, 395, 663 
\bibitem[Landau \& Lifshitz(1987)]{Landau} Landau, L. D. \& Lifshitz, E. M. 1987,
    Fluid Mechanics, 2nd ed. Pergamon
\bibitem[Mart\' i \& M\" uller(1994)]{Marti}Mart\' i,J.M. \& M\" uller,E. 1994, J.Fluid Mech., 258, 317
\bibitem[Matzner \& McKee(1999)]{Matzner}Matzner,C.D. \& McKee,C.F. 1999, \apj , 510, 379
\bibitem[M\'esz\'aros \& Rees(1997)]{Meszaros} M\'esz\'aros, P.~\& Rees, 
M.~J.\ 1997, \apj, 476, 232 
\bibitem[Nakamura \& Shigeyama(2004)]{Nakamura} Nakamura, K.~\& 
Shigeyama, T.\ 2004, \apj, 610, 888 
\bibitem[Perna  \& Vietri(2002)]{Perna}Perna,R. \& Vietri, M. 2002, \apj , 569, L47
\bibitem[Rees \& M\'esz\'aros(1992)]{Rees} Rees, M.~J.~\& 
M\'esz\'aros, P.\ 1992, \mnras, 258, 41P 
\bibitem[Rhoads(1997)]{Rhoads} Rhoads, J.~E.\ 1997, \apjl, 487, L1 
\bibitem[Sakurai(1960)]{Sakurai} Sakurai, A. 1960, Commun.Pure Appl.Math.,13,353
\bibitem[Tan, Matzner \& McKee(2001)]{Tan}Tan, J. C., Matzner, C. D. \& McKee, C. F. 2001,\apj, 551, 946
\bibitem[van Paradijs et al.(1997)]{van} van Paradijs, J., 
et al.\ 1997, \nat, 386, 686 
\bibitem[Fields et al.(2002)]{Fields2} 
\end{thebibliography}
\end{document}